\documentclass{article}
\usepackage[english]{babel}
\usepackage[letterpaper,top=3cm,bottom=3cm,left=3cm,right=3cm,marginparwidth=1.75cm]{geometry}

\usepackage{amsmath}
\usepackage{graphicx}
\usepackage{color}
\usepackage[colorlinks=true, allcolors=blue]{hyperref}
\usepackage{authblk}

\usepackage{enumitem}

\usepackage{xspace}

\usepackage[sorting=none,style=numeric-comp]{biblatex}
\begin{document}

\begin{titlepage}

\vspace*{-1.5cm}
\centerline{\large THE HEP SOFTWARE FOUNDATION (HSF)}
\vspace*{1.5cm}
\noindent
\begin{tabular*}{\linewidth}{lc@{\extracolsep{\fill}}r@{\extracolsep{0pt}}}

\\
 & & HSF-TN-2024-01 \\  
 & & April 2024 \\ 
 & & \\
\end{tabular*}

\vspace*{.5cm}

{\bf\boldmath\huge
\begin{center}
  Analysis Facilities White Paper
\end{center}
}

\vspace*{1.0cm}

\begin{center}


D. Ciangottini$^{1, \rm b}$,
A. Forti$^{2, \rm b}$,
L. Heinrich$^{3, \rm b}$,
N. Skidmore$^{4, \rm b}$, 

C. Alpigiani$^{5}$,
M. Aly,$^{2}$,
D. Benjamin$^{6}$,
B. Bockelman$^{7}$,
L. Bryant$^{8}$,
J. Catmore$^{9}$,
M. D'Alfonso$^{35}$,
A. Delgado Peris$^{17}$,
C. Doglioni$^{2}$,
G. Duckeck$^{10}$,
P. Elmer$^{11}$,
J. Eschle$^{12}$,
M. Feickert$^{13}$,
J. Frost$^{14}$,
R. Gardner$^{8}$,
V. Garonne$^{6}$,
M. Giffels$^{36}$,
J. Gooding$^{15}$,
E. Gramstad$^{9}$,
L. Gray$^{16}$,
B. Hegner$^{31}$,
A. Held$^{13}$,
J. Hernández$^{17}$,
B. Holzman$^{16}$,
F. Hu$^{8}$,
B. K. Jashal$^{18,19}$,
D. Kondratyev$^{20}$,
E. Kourlitis$^{3}$,
L. Kreczko$^{21}$,
I. Krommydas$^{22}$,
T. Kuhr$^{10}$,
E. Lancon$^{6}$,
C. Lange$^{23}$,
D. Lange$^{11}$, 
J. Lange$^{34}$, 
P. Lenzi$^{1}$,
T. Linden$^{24}$,
V. Martinez Outschoorn$^{27}$,
S. McKee$^{25}$,
J. F. Molina$^{17}$,
M. Neubauer$^{26}$,
A. Novak$^{35}$,
I. Osborne$^{11}$,
F. Ould-Saada$^{9}$,
A. P. Pages$^{28}$,
K. Pedro$^{16}$,
A. Perez-Calero Yzquierdo$^{17}$,
S. Piperov$^{20}$,
J. Pivarski$^{11}$,
E. Rodrigues$^{29}$,
N. Sahoo$^{30}$,
A. Sciaba$^{31}$,
M. Schulz$^{31}$,
L. Sexton-Kennedy$^{16}$,
O. Shadura$^{32}$,
T. \v{S}imko$^{31}$,
N. Smith$^{16}$,
D. Spiga$^{1}$,
G. Stark$^{33}$,
G. Stewart$^{31}$,
I. Vukotic$^{8}$,
G. Watts$^{5}$,

\bigskip
{\it\footnotesize
$^b$Editor,
$^1$INFN,
$^2$University of Manchester,
$^3$Technische Universität München,
$^4$University of Warwick
$^5$University of Washington
$^6$Brookhaven National Laboratory
$^{7}$Morgridge Institute for Research
$^{8}$University of Chicago
$^{9}$University of Oslo
$^{10}$Ludwigs-Maximilians Universität München
$^{11}$Princeton University
$^{12}$Syracuse University
$^{13}$University of Wisconsin-Madison
$^{14}$University of Oxford
$^{15}$Technische Universität Dortmund
$^{16}$FNAL
$^{17}$CIEMAT
$^{18}$IFIC
$^{19}$TIFR
$^{20}$Purdue University
$^{21}$University of Bristol
$^{22}$Rice University
$^{23}$Paul Scherrer Institute
$^{24}$Helsinki Institute of Physics
$^{25}$University of Michigan
$^{26}$University of Illinois at Urbana-Champaign
$^{27}$University of Massachusetts Amherst
$^{28}$IFAE
$^{29}$University of Liverpool
$^{30}$University of Birmingham
$^{31}$CERN
$^{32}$University of Nebraska-Lincoln
$^{33}$SCIPP UC Santa Cruz
$^{34}$University of Hamburg
$^{35}$Massachussets Institute of Technology
$^{36}$Karlsruhe Institute of Technology
}
\end{center}

\vspace*{1.0cm}

\begin{abstract}
  \noindent

This white paper presents the current status of the R\&D for Analysis Facilities (AFs) and attempts to summarize the views on the future direction of these facilities. These views have been collected through the High Energy Physics (HEP) Software Foundation’s (HSF) Analysis Facilities forum~\cite{hsfaff}, established in March 2022, the Analysis Ecosystems II workshop~\cite{ecosystemstwo}, that took place in May 2022, and the WLCG/HSF pre-CHEP workshop~\cite{prechep}, that took place in May 2023. The paper attempts to cover all the aspects of an analysis facility.

\end{abstract}

\vspace{\fill}

{\footnotesize
\centerline{\copyright~Licence \href{http://creativecommons.org/licenses/by/4.0/}{CC-BY-4.0}.}}
\vspace*{2mm}

\vskip 0.5cm
\textbf{Keywords}: High energy physics, analysis facilities, data analysis, scientific computing, data access, grid computing, federated identity management, analysis preservation, resource provisioning, HL-LHC

\end{titlepage}

\clearpage

\tableofcontents

\section{Introduction}\label{introduction}

In the HL-LHC era (to begin in 2029) LHC analysts, depending on the type of analysis they do, may have to process up to an order of magnitude more data than currently. At the same time, there are significant changes to analysis techniques such as the introduction of columnar analysis, parallelised data access and the organization of multi-step workflows into pipelines, some of which may be better suited to run using cloud technologies and heterogeneous resources. Consequently, the analysis infrastructure must evolve. This is an active area of R\&D collectively referred to as “Analysis Facilities”. 

An Analysis Facility (from here on abbreviated to AF) can be defined as infrastructure and services that provide integrated data, software and computational resources to execute one or more elements of an analysis workflow. These resources are shared among members of a Virtual Organization (VO) and supported by the organization. This R\&D is expected to be integrated with the existing infrastructure and guide any future development.

Since discussions on the evolution of analysis for the HL-LHC started, the analysis infrastructure has often been considered as something separate from the current grid infrastructure, mainly due to two aspects: large scale interactivity for fast analysis cycles and integration of cloud native technologies. The possible additional requirement to support whole VOs also sets it apart from the local Tier 3 sites currently giving access only to local users and occasionally to external collaborators. The nature of AFs has provoked long debates. CERN satisfies the requirements described above but it is the only site supporting all users and the competition for resources there is already quite intense. The question is how to define an AF that offers similar functionalities as those at CERN without hindering participation from other sites. Can they support a subset of functionalities or be open to groups of users and still be considered an AF?

We can start by stating that the most common setup for analysis (small interactive machines, a batch system and a shared file system between the two) will not be replaced and we should focus on a technology refresh. The effort is to complement and add new tools to what already exists.  A second observation is that the distinction between AFs and the Worldwide LHC Computing Grid (WLCG)~\cite{wlcg} is more blurred than currently appreciated and the integration of AFs within existing WLCG infrastructure is important for broad accessibility and a standardized user experience with smooth data access, dynamic allocation of WLCG computing as part of the AF resources and easier support from the sites. A third observation is that several analysis facility sites support multiple experiments, for instance CERN, FNAL and DESY, and the aim is to provide the same infrastructure for most if not all experiments. In other words AFs do not exist in a vacuum and should not be considered isolated and one should not assume the resources are dedicated. The way forward is not to focus on a specific architecture, but on the building blocks that can be deployed whatever shape an AF takes. For these reasons we should talk about “analysis infrastructure” more than “analysis facilities”.

In this spirit, following the Analysis Ecosystem II workshop and the Analysis Facilities Forum presentations, a number of key areas were identified which this white paper addresses:

\begin{itemize}
    \item \textbf{Users' perspective and their use cases} (Section~\ref{userperspective}): this should drive the R\&D and naturally will be discussed first.
    \item \textbf{Compute resources access and provisioning} (Section~\ref{computeresources}): considers the evolution of computing requirements, in particular the concept of scale out resources and integration on existing infrastructures.
    \item \textbf{Data Organisation, Management, Access (DOMA)} (Section~\ref{doma}): concerns the input/output data for analysis workflows. In particular the need to integrate AFs with grid storage and methods to solve the data locality problem. 
    \item \textbf{Federated Identity Management and and Authentication, Authorization Infrastructure (AAI)} (Section~\ref{aai}): a consistent AAI across analysis infrastructure is at the core of a smooth integration particularly for data access; federated identity can also allow access to multiple facilities. 
    \item \textbf{Accelerator resources} (Section~\ref{accelerators}): considers the increased use of accelerators in analysis and the resulting need to add accelerators to AF resources.
    \item \textbf{Analysis portability and preservation}(Section~\ref{preservation}): handles the requirements on the packaging of software to enable collaborative analysis across facilities with the ability to rerun or recast analyses in the future. 
    \item \textbf{Monitoring and Metrics} (Section~\ref{monitoring}): concerns the need for appropriate monitoring and benchmarking both for users to effectively use resources and for sites to provide resources more effectively.
    \item \textbf{End user documentation and support} (Section~\ref{documentation}): expresses the requirement for  findable, quality documentation and support for users as part of the AF infrastructure is essential.

\end{itemize}

This White Paper does not aim to answer all the questions surrounding analysis facilities and infrastructures. We have identified remaining, key open questions for the community and they are highlighted at the end of the relevant sections.

\section{User perspective}
\label{userperspective}

Analysis Facilities aim to facilitate the preparation and execution of HEP data analyses. During the course of the HSF Analysis Facility Forum a number of requirements have been identified in order to enable users to do their research effectively. 

\subsection*{Ability to perform fast research iterations on large datasets interactively}

Research success relies crucially on the ability to test ideas during the R\&D phase of an analysis, e.g. assessing the impact of a change in event selection or determining the impact of a variation in systematics. Here a fast “time-to-insight”, i.e. minimal wall-clock time, is a priority. In this phase, the user expects to rapidly gain access to and interactively work with the resources. This requirement so far has been satisfied by classic facilities that provide a number of interactive, single “login” nodes accessible via remote terminal/ssh sessions or, more recently, by providing small single node Jupyter~\cite{jupyter} notebooks.

However dataset sizes are increasing and an interactive experience constrained to a single node with a handful of resources may no longer be sufficient. The requirement becomes to enable a distributed and interactive analysis style, where the work is coordinated by the user in an interactive session, but the user can dynamically connect to a scale-out system that provides additional resources. The expectation is that the processing on additional resources starts in parallel almost immediately, with only a short lag-time due to resource provisioning, and the resulting data is aggregated and sent back to the interactive node for the user to examine in real-time. 

What constitutes “minimal wall-clock time”, “large amount of resources” and “large datasets” is analysis dependent and experiments should define acceptable/expected ranges for their analyses. The implications on how interactivity could be achieved are still under discussion. 

\subsection*{Ability to convert interactive to batch-schedulable workloads}

When an analysis matures from an R\&D phase to performing the final measurement, the usage pattern often changes from a rapid iteration and frequent changes of the analysis to a more controlled pattern, in which results from various configurations must be produced. Here users may not urgently require full interactivity and, in fact, may prefer to prepare a number of analysis configurations ahead of time and asynchronously schedule them in bulk. In this mode, the total wall-clock time required for a processing is not an urgent concern to the user and facility-side optimizations towards resource use efficiency, load balancing, may be valid trade-offs. This mode is typically handled as a non-interactive batch workload, as supported by a workload scheduling system.

As more interactive distributed workloads become possible for the R\&D phase, it is important that it is easy to convert a workload from an interactive setup into a “headless” mode that is amenable to batch systems, this poses requirements on both the analysis software as well as the analysis facility. This of course is mostly important in case of Jupyter notebooks where the submission procedures to a batch system need to be developed by the AFs. A Jupyter notebook that, in interactive mode, dynamically scales to multiple nodes to execute an analysis and is driven by a user working within a notebook, should be re-configurable into a non-interactive workload that requests a fixed number of workers, when they become available, and scales down workers reliably to free resources again.

\subsection*{Ability to interact with the WLCG and scale outside of the facility on occasion}

Given the distributed nature of HEP research and the existence of the WLCG, analyzers often face the need to interact with the wider infrastructure. Common tasks may include interacting with other WLCG and other opportunistic resources, pulling or pushing data in and out of the facility. As such, AFs should not be “sealed environments”, but integrated into the wider WLCG infrastructure.  If the workload would exceed traditional analysis activities and/or the data is distributed across the WLCG and the task at hand matches well to grid-style processing the user should be able to use grid middle-ware tools to offload the work to the WLCG and connect to opportunistic resources. The latter may be particularly important for training large machine learning models for which AFs may not have sufficient GPUs. The ability to convert interactive workflows to batch-like jobs (as described above) should also help with this use case.

\subsection*{Ability to efficiently train machine learning models for HEP}

The progress over the last decade in Machine Learning (ML) techniques invariably also affects physics analysis in HEP, where it is used across the board in data-taking, simulation, reconstruction and analysis. ML methods play a dual role: they can be used to find more powerful algorithms, but also enable significant acceleration due to the high degree of parallelization possible in neural networks, which maps well to hardware such as GPUs. For AFs this implies two important requirements: 1) the need for an efficient python-based analysis environment and 2) access to heterogeneous hardware both interactively as well as in batch mode. The first is due to the dominance of python within ML research with tools like PyTorch~\cite{pytorch}, SciKit-Learn~\cite{scikit-learn}, and Tensorflow~\cite{tensorflow}, which is expected to continue. ML development also requires early on data- and compute-intensive fast iteration during the development of the model, where interactive access to, e.g. Jupyter notebooks is a priority. The second requirement of accelerator availability must be met both interactively as well as non-interactively. As a user, it is expected to be able to start an interactive session with GPU access. For a positive user experience it may be more important to be able to quickly access a single GPU than to be able to keep it for a long time or scale from a single GPU to a multi-GPU environment. Access to batch GPUs is increasingly required for model selection and hyperparameter tunings but also for cutting-edge ML research, because both models and datasets are growing, and a typical training run for many models now exceeds 24 hours. Here the user should be able to use standard workload schedulers to describe the shape of the workload (e.g. GPUs, CPUs, memory). 

\subsection*{Ability to reproducibly instantiate desired software stack}

Analyses are a mixture of both collaboration-wide software frameworks and more custom-tailored packages that suit the particular analysis needs of the team. As a user, it is reasonable to expect the ability to assemble this mixture. As an additional aspect, the ability to archive and possibly distribute this particular software environment is important. This may be of particular interest for both analysis preservation and reuse purposes, but also help towards the ability to reinstantiate the analysis environment at different facilities. CVMFS~\cite{cvmfs} --- the dominant solution in HEP for software distribution --- as well as linux containers are expected to contribute here. Users may not be experts in these technologies and thus the process of listing requirements for the software environment may need to be abstracted from the underlying technology.

\subsection*{Ability to collaborate in a multi-organisational team on a single resource}

Analysis teams at HEP experiments are multi-organisational: team members may not be at the same institution, region or even country, yet the goal is to jointly work towards a physics result. Analysis is also distributed: responsibility for parts of the analysis workflow is often taken on by a subgroup and the resulting data products of one subgroup must be seamlessly passed onto another one. Such a requirement is easiest to fulfill if access to resources are equitable, where the full team can work with the same resources. The main reason for this is that maintaining copies of data, i.e. downloading datasets, ensuring all files are present, doing necessary post-processing/combining/bookkeeping on the local files before they're ready for analysis is hard work for the users, inefficient and error-prone. Having one copy means this only needs to be done once. In the current model this is ntuples, but this could also refer to skims of reduced formats in the future. 

Given the data volumes of an analysis it may be desirable to work on the same resource for the duration of an analysis project, however, fine-grained control over which resource this would be may not be necessary and may be left up to the VO, i.e. the VO may assign users or physics groups to AFs rather than users flocking to a facility or another. 

\subsection*{Ability to move analyses to new facilities}

While it should be possible for the full analysis team to access the same facility for jointly developing the analysis, users and/or system administrators may value the ability to move an analysis effort between facilities. This implies the requirement that the data products that are local to a single facility can be replicated to distributed storage. Note that this does not imply a requirement for a distributed, fast, automatic synchronization but rather an on-demand method of data movement. 

\subsection*{Ability to efficiently access collaboration data as well as make intermediate data products available to the team}

Analysis starts after a common bulk pre-processing of the data by the collaboration on WLCG resources. It is therefore necessary to be able to interact from within the facility with distributed data management services, such as Rucio~\cite{rucio}, both from interactive and non-interactive environments. Notably, the physical presence of the data on the processing site is not a priority of the user as long as a smooth processing of the data can be ensured on-demand. This opens the possibility of inbound data-access via caching services such as XCache~\cite{xcache}, where the user only references a global identifier and the file is made available on-demand.

During user data processing new data products are being created which need to be made available within the facilities’ distributed storage. Additionally it is important that users are able to share data to other collaborators (possibly working on other facilities). Currently this is often satisfied through an ad-hoc choice of a storage solution that many collaborators have access to (e.g. EOS~\cite{eos}) as a “data exchange”. Ideally, however, it would be possible to register user data seamlessly in a global namespace. It may be sufficient that such data publishing is based on explicit intent and full POSIX semantics may not be needed. Thus an implementation of such a requirement does not imply a global, automatically synchronized filesystem (such as AFS that provides networked file storage for CERN users).
 
\subsection*{Ability to express interdependent distributed computations at small and large scales}

An analysis at LHC scale is often the result of a complex workflow of multiple interdependent steps that need to be executed in a certain order: for example data reduction/event selection, followed by preparation of the inputs required for statistical analysis, followed by inference and visualization steps. Here, it is often the case that due to the distributed nature of analysis teams, steps are largely independently developed by analysis subgroups and later on integrated into an overall data analysis pipeline. The individual steps may have different, and even conflicting, software environment requirements. This implies users do not just need to set up one environment, but ideally seamlessly transition between many. Nonetheless, it is desirable for analyzers to be able to express and schedule the full data analysis pipeline into an overall workflow. 

Workflows in this sense should be contrasted to computational graphs that are, e.g. built during the use of a distributed data processor, such as Dask~\cite{dask}, as each node in the former may be a full analysis step (e.g., running the fit), while the latter typically considers nodes to be small computational units (e.g., compute invariant mass). Such workflows have been used both for analysis reinterpretation work, where the full analysis pipeline is re-executed on new simulation data, as well as for automation of, e.g. the production of calibration functions for analysis. 

\subsection*{Open Questions}

\subsubsection*{On interactivity of analysis sessions}
What is the user expectation of an interactive session? Interactive analysis sessions on large datasets implies the ability to scale out to distributed resources and users will expect instant access to such an analysis session. What are the required disk resources, computing cores (order 100s) and network capabilities to realize this? The answers to these questions are analysis/workflow dependent, vary widely, and will need to be quantified as such. (The definition of interactive remains an open question for the community). The question is very pertinent to analysis in the HL-LHC era, where datasets will be much larger (~10x) for ATLAS and CMS from Run 4 onwards and will also grow again for LHCb and ALICE from Run 5. This does not automatically mean that interactive needs jump by 10x, but an increase will happen and should be quantified.

\subsubsection*{On the Interfaces to resources provided}
Analysis work is multi-faceted and while a browser-based interface can be attractive for some use-cases such as interactive computing with notebooks, it should not be the only possible interface to the resource. Should users expect to be able to connect via their preferred text-based terminals or Integrated Development Environments (IDEs) such as Microsoft's Visual Studio Code?

\section{Compute resources: access and provisioning}
\label{computeresources}

End-user data analysis workflows are evolving significantly, with new data formats designed by the LHC collaborations to reduce the storage footprint and be more compatible with modern data science paradigms, eventually enabling researchers to utilize frameworks that require little to no experiment-specific software. This evolution has, inevitably, an impact on the infrastructure and resources dedicated to user analysis activity.

Outlined are the current R\&D activities that have been reported to the HSF dedicated forum, with particular focus on the evolution of access and provisioning of the compute resources to expose heterogeneous and possibly distributed resources. A set of recurrent questions and discussions are the starting point for the review targeted by this section: 
\begin{enumerate}
    \item \textbf{Compute resources access:}
    \begin{enumerate}[label=(\alph*)]
        \item Workflow management: tools leveraging batch systems in a native way are arguably the most widely adopted ones at the time of writing, nevertheless workflows are becoming larger and more complicated and the integration of cloud paradigms for a seamless transition from local “interactive” development to full scale analysis multi-step execution are emerging in many R\&D activities. That is expected to enable enhanced data exploration and reduce “time-to-insight”.
        \item UX/UI: ssh-ing into a preconfigured machine is a de-facto standard for most analysis and that is not expected to change. Instead, a possible enhancement of the UX is expected to come from WebUI interfaces largely adopted in the context of cloud based data science ecosystems.
        \item 
        AAI: the adoption of tokens for Authentication and Authorization by the HEP community is also important in the context of AFs and integration of cloud technologies as well as simplified access to data from diverse resources. Section~\ref{aai} is dedicated to AAI infrastructure.
        \item Containers: a base hypothesis of the computing models under investigation is that the payloads are containerized. Section~\ref{preservation} details the portability and reproducibility of that approach.
    \end{enumerate}
    \item \textbf{Compute resources provisioning:}
    \begin{enumerate}[label=(\alph*)]
        \item Provisioning responsiveness: point 1.a shows a significant shift in terms of the latency that is acceptable in the interactive vs. non-interactive scenarios. That opens the question of how to implement a computing model at AF level where dynamic provisioning can possibly live alongside crowded traditional batch systems.
        \item Efficiency and performance: upcoming frameworks that are enabling a whole set of new possibilities from the user perspective have to be validated against their efficiency in terms of resource utilization and overall physics throughput. 
        \item Site provisioning model: HPC centers and commercial clouds are going to have a significant impact on the future of resource provisioning. Finding a way to integrate them seamlessly is crucial, along with the understanding of which kind of analysis paradigm can benefit the most from that integration. 
        \item Resource provisioning model: the server level is the other side of the spectrum of the provisioning model. One solution could be to allocate whole nodes and reserve a whole “fat” node that is fully committed to achieve the quickest analysis turnaround time. It remains to be seen if this can be sustainable in the future and compatible with the goal of maximum overall resource efficiency.
        \item Specialized hardware: the adoption of GPUs and, potentially, FPGA acceleration, is expected to increase, and this has to be made both easy and efficient from the infrastructure perspective. Although not many case studies have been reported, we do expect this to be a strategic point to investigate. This is transversal to both access and provisioning patterns. 
    \end{enumerate}
\end{enumerate}

Many R\&Ds reported to the AF Forum on their investigations, which allows an “educated guess” on what the main traits are for future analysis infrastructures. 

\subsection*{User Interfaces}

The SSH login into a pre-configured UI still holds as a requirement for many use cases, especially for non-pythonic workflows like scripting for batch job submission. For analysis of reduced data (at the moment often done in the previous step), an interactive interface can be accessed through the adoption of tools enabling a single WebUI to log in and access the resources they need. JupyterHub is a popular choice for this, because it allows multiple users to spawn JupyterLab instances on a cluster of servers. One interesting model emerged for point 2.a: providing users with a relatively small amount of “always-on” resources, ready to get the user started with an interactive workflow, while scheduling a larger fraction on lazier batch system queues, leveraging offloading mechanisms (see below). Another area of investigation is the possibility of using JupyterHub as a mechanism for ssh-UI on-demand provisioning, integrating the SSH access into a running JupyterLab instance, thus enabling users to share a common environment with access to different UIs.

\subsection*{Offloading to external resources}
\begin{figure}
    \centering
    \includegraphics[height=6cm]{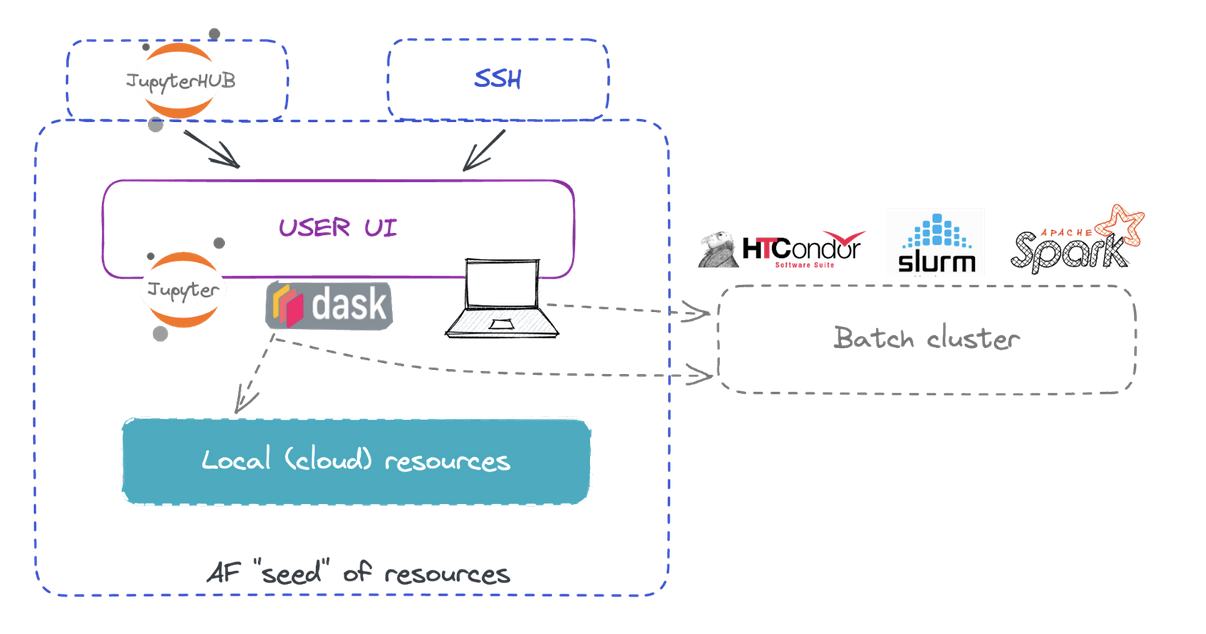}
    \caption{High level schema of the recurrent patterns reported at the forum by R\&D activities around compute access and provisioning models evolution.}
    \label{fig:AFschema}
\end{figure}

In terms of supporting the offloading of “heavy lifting” workflows to external resources, and since the possibility to distribute batch jobs is still a requirement, most of the reported activities aim to integrate a mechanism for leveraging the batch systems as an offloading backend for the more “interactive”/data-science-like analysis workflows. That behavior has been demonstrated mostly via Dask cluster deployments. In fact, many of the presented solutions have their foundation in Dask distributing payloads on either HTCondor~\cite{htcondor}, SLURM~\cite{slurm}. Some sites have deployed Apache Spark~\cite{spark} managed clusters, notably CERN Analytix cluster. For the requirements of fast prototyping of ML models and even full analyses, an offloading mechanism for a whole JupyterLab instance to a remote, specialized node that is fully committed to the analysis for a short period of time could be considered. Although possible, it has yet to be proven sustainable at scale with dedicated activities. With all of these considerations in mind,  figure~\ref{fig:AFschema} attempts to schematize a high level description of an AF resource management “strawman”. 

Different models for the evolution of the access and provisioning of analysis resources have been presented by several groups (SWAN~\cite{swan}, Coffea Casa~\cite{coffea}, NAF~\cite{NAF}). Most of the commonalities are in the methods of providing access to the users’ UIs. Providing SSH and/or JupyterHub access is becoming a default mode of operations in many (if not all) the presented activities. Interesting variations occur for the scale-out mechanism toward remote resources. A number of facilities are satisfying scenarios where the batch resources are co-located (in terms of network segregation) with the AF interactive seed, differing only in the integration with the analysis tools. On a slightly different path, there is an evaluation for integrating geo-distributed centers (e.g.,T2 grid resources) as offload candidates for tasks (INFN~\cite{INFN_distributed}); this presents the problem of having the JupyterLab instance on a different network from the worker nodes. The current solution in such a scenario is the creation of a self-contained cluster to be created at the remote site on-demand. 

One final observation is that there are ongoing efforts, presented by the kubernetes~\cite{k8s} user activities, that might allow for an alternative implementation. Here, payloads and their distribution can eventually be fully delegated to kubernetes native resources with the introduction of dedicated abstractions and scaling improvements. There are preliminary implementations and R\&D efforts that have been reported at the forum but have yet to be investigated properly in order to make a clear comparison with the alternatives presented above.

To support complicated workflow pipelines, as expressed in the user requirements, workflow languages such as Snakemake~\cite{snakemake}, Yadage~\cite{yadage}, Common Workflow Language (CWL)~\cite{cwl} and others may be suitable. Also, some workload schedulers have features that allow users to express dependencies between workloads (e.g. HTCondor's tool DAGMan). Workflow services such as REANA~\cite{reana}, capable of processing such languages, should be integrated into the existing batch infrastructure at the facility in order to have access to the same data and compute resources that would be available if the pipeline were processed manually. Note that REANA also works with a kubernetes back end. 

The large number of new technologies may trigger the question: should users be exposed to all of these or should there be a common entry point interface? Reaching agreement on a common interface will be challenging, the introduction of Jupyter Notebooks/JupyterLab is an attempt at having a common interface but this remains an open question. 

\subsection*{Open Questions}

\subsubsection*{On resource interoperability?} If we cannot guarantee uniformity of technologies how do we manage “interoperability” between resource providers?

\subsubsection*{On a common interface}
Given the large number of new technologies should users be exposed to all of these or should there be a common entry point interface such as JupyterLab?

\section{DOMA (Data Organisation, Management, Access) }
\label{doma}

Achieving the transparent movement of data between AFs and other storage systems used within HEP is still a matter of development. Fast access to input data is one of the most important aspects of an AF. Considerable effort is going into reducing the input data volume handled by end users so that it can fit comfortably at single sites. However, not all analyses will be able to use such “reduced” formats and whilst the site may have the storage space, it may not have the interactive computing resources for an AF (for example a T1 site). A second aspect that needs to be taken into consideration is where users’ output will be stored and if it needs to be shared. Experiment analysis models are expected to develop such that direct large-scale ntuple productions by analysis groups are very uncommon. The new reduced formats are expected to replace these,  with large productions continuing to be carried out on the grid. 

\subsection*{Data access: single site, replicated or cached}

Users need to be able to find their input data as they would on any other type of experiment resource and be able to access them in a timely manner. There are a few possibilities to organize the data and these shape the AF infrastructure:

\begin{itemize}
    \item Put everything at CERN. This of course is possible but the interactive and batch system resources should also increase to cope as well as the user support response to the increased requests.
    \item Create other facilities that host  a large storage common to all computing resources and a large amount of interactive/batch resources, similar to those found at CERN. Typically this would be at T1s or large labs like DESY. 
    \item Remote access to all the data from interactive computing resources with the latency hidden by the caches. This already exists at some sites like UCSD and NERSC.
\end{itemize}

The last of these models can maximize the usage of any type of available computing resource and does not stop either of the previous models from being implemented.

Analysis workflows frequently access the same data repeatedly, so caching is beneficial. So far all testing has been done using XRootD~\cite{xrootd} based XCaches~\cite{xcache}, which have two advantages: the first is that the applications (including uproot~\cite{uproot} and RData Frame\cite{rdataframe}) can read only the required portions of files, reducing the volume of transfers; the second is that they are already used at HEP sites so it just matter of ironing out the more specific user requirements. Hosting managed copies of the data is usually painful, unless a grid storage system is available and integrated in the experiment’s system. Caches, however, require little maintenance. If the deployment of a cache is simple enough that it can be adapted to be used at Tier3 sites then these resources could continue to be used effectively without hosting managed data. This is particularly important considering the analysts will want local user formats (nanoAOD/PHYSLITE) skimmed (a typical case is  multi-lepton analysis) or augmented (i.e. custom objects that need tracks or particle flow candidates). Augmented files may point to external remote files, so the ability to copy locally the augmentation onto a cache is almost mandatory to reduce latency. There is the possibility to use https for caches, but that is not as well tested and would require further R\&D. Using caches means the HL-LHC data, in whatever format, has to be on remotely accessible storage that can be integrated with the caches, i.e., it cannot be on purely internal storage. There is also a question on the specifications of these caches. They need to be fast, e.g. on SSD and NVMe, and possibly support a large number of users; but they could also be large HDD which are not as fast but store a lot more data. For example the Southern California (SoCal) Cache is a ~2PB cache serving CMS analysis jobs from both UCSD and Caltech and is made out of retired HDDs. The scalability comes from using several data servers grouped under a redirector (similar to a regular XRootD cluster). 

\subsection*{Common namespaces} 

Depending on the preference of the experiment on how to track data access, access to remote data may benefit from integrating the data management tools into the user interface. Usually users query their experiment’s DDM (Distributed Data Management) system  (e.g. Rucio or Dirac~\cite{dirac}) from the command line, but considering the request of interactive resource access via IDE, it is interesting to highlight the integration of Rucio with a Jupyter notebook plug-in. This was developed by Rucio and the ESCAPE~\cite{escape} project and avoids a double step of having to query Rucio in a terminal and then upload the resulting list in the notebook. The plug-in could have configurable caches available. In this case the DDM tools provide the data common namespace. It is also possible to access a common namespace using a combination of XRootD redirectors and caches like CMS does, however here the user still has to know the list of files they want to access, requiring interaction with a file catalog. 

Another example of a common namespace is when site storage underpins all its computing resources. Usually this is a POSIX-like file system. The most prominent example for this is EOS at CERN. EOS offers a POSIX-like file system, with a single namespace, that can be accessed by interactive resources, local batch and cloud resources. Grid jobs and remote non-grid resources can also read and write data via other protocols and the users can always find them easily via the POSIX-like interface. This setup is one of the reasons CERN is so popular for analysis --- everything is easily accessible with a variety of different tools.

\subsection*{POSIX vs Object stores}

Full POSIX compliance cannot be expected and is also very unlikely to be necessary  for complexity and for performance reasons. In this text, POSIX should be implicitly understood as POSIX-like.

Users favor POSIX file systems because they are more familiar and intuitive and the experiments’ applications and services are also built on POSIX. The emergence of object stores, with their scalability and efficiency in serving data, raises the possibility that new workflows can be adapted to use distributed object stores. Scalability of hierarchical POSIX file systems is a recurring worry, and further effort to understand if data access speed and scalability are limited by application logic or storage technologies is needed. One should also clearly distinguish between the use cases: accessing several TB of data from mass storage vs copying a few smaller files in a local directory. One needs to take into consideration the evolution of analysis software, like ROOT. Traditionally it is bound to local IO and relies on POSIX semantics, which may use "advanced" concepts (symlinks, hardlinks, fseek()...) for maximum performance and convenience. Some analysis software may already provide data IO interfaces for simpler protocols (such as Amazon's S3 and XRootD). However even though the tools provide those interfaces, users usually wrap their analysis in their own code that will use a local file system, because that is what they are familiar with. For example, a user may create a script to chunk a data file into smaller files relying on local Unix tools rather than learning how to do that in S3. Users may also want to explore a new dataset or download some data to their local computers in a format with which they are familiar, which is static files with POSIX semantics for access/interaction. This happens on a small scale: users are not interactively exploring or exporting terabytes to petabytes of data. Therefore, the only requirement is that an object store can export (usually small amounts of) selected data to a static file. Similarly, user output (histograms, etc.) from analysis workflows is orders of magnitude smaller and may be input to subsequent tasks that are separate from the AF and any object store semantics. Therefore, the ability to output static files is useful, but separate from storing, accessing, and caching TB-PB of data. "Skimming", keeping only selected subset of events for further analysis, is an intermediate case, and one probably wants the option to output into an object store format for this case. Taken together, all of this implies a general need for both static files and object stores to be accepted as both input and output formats, with the most efficient formats preferred for the largest-scale processing tasks. 

There have been a few developments in applications looking to support object stores like the ROOT I/O system RNtuple~\cite{rntuple} or XRootD framework. However a wider shift to object stores would require a dedicated working group to assess the pros and cons of integrating them into experiment software. In the future it could be possible to access large scale object stores directly from the analysis software while continuing to support POSIX-like features for smaller more interactive tasks. 

Whatever architecture an AF may eventually have (distributed, replicated at T1s, integrated with national resources) a storage system that is accessible from everywhere and can be easily replicated should be the aim. Different sites may have their own solutions, and often a different type of storage for different resources with different protocols, this poses a challenge for users to migrate to other facilities or share their data. Currently this role is filled by the grid storage which is not designed for user experience, particularly without going through the experiments’ DDM.

\subsection*{Integration of namespaces and access methods}

As expressed in the user requirements, users want the ability to move from one resource to another. This implies the ability to share the data as well as the code with colleagues. For this   a possible solution may be for facilities to offer both local storage, that computer nodes can read and write to and from efficiently, as well as storage whose contents will be registered in the wider DDM infrastructure. While DDM tools such as Rucio support user uploads, their current main use is the upload of output data from WLCG jobs, so it would be necessary to extend this to analysis facility output datasets. 

For many analysis activities users currently use file systems available at well-known sites, such as EOS, to share data. Within the HSF AF Forum the idea of a global and shareable working space, has been discussed. However, it would be necessary for such infrastructure to be extendable to GB-TB scale datasets that may be produced during analysis. There is no obvious scalable and performant solution here. We cannot resort to distributed file systems like AFS for performance reasons, sharing services like CERNbox~\cite{cernbox} are not suited for large data sharing  and “common spaces” on grid storages need easier interfaces for the users.

There are all sorts of details to be ironed out to support storage systems with different namespaces and access methods. Whether talking about POSIX vs object stores, grid storage vs local storage, different access methods at the same site or integrating a WEB based service, like CERNbox, the problems are similar. It will require work to map each solution to the access of another, for example how CERN integrates SSO and Kerberos for Binder~\cite{binder}. Some solutions may work only for fuse mountable protocols (EOS/XRootD) and not for filesystems with kernel drivers. There is also the option adopted by LHCb to encourage users to use their Analysis Production Data package (apd)~\cite{apd} tools to find data also locally which ATLAS and CMS could do using Rucio.   Alternatively, moving away from POSIX has been proposed, but this clashes with the fact that POSIX is at the base of any Linux server and to avoid it we should forbid users from interacting with file systems, which is not feasible. Going forward if more cloud technologies are introduced it will be easier to integrate object stores from the AAI point of view but also from the infrastructure point of view. For example, whereas a broken POSIX/fuse mount is very intrusive to workloads, object stores are more loosely coupled and make it easier to recover from issues with the backends. If we look at the integration with containers, POSIX and fuse in containers often require additional components and even additional privileges. Object stores, being native cloud storage, integrate very well with user containers. 

\subsection*{CERNBox data and document sharing}

File sharing services like CERNBox, heavily used for data and document sharing, can benefit from the fact that the output data of jobs run at the CERN facility can be written on a dedicated EOS space, making them automatically available to colleagues. CERNBox has other functionalities, such as integration with Zenodo that makes it really interesting as a document sharing service. Another useful workflow that CERNBox enables is to combine online work at the AF with offline work in a personal computer, and have the files synchronized via the CERNBox sync client installed on the machine. The CERNbox team is planning to also extend support beyond EOS to CephFS which will make the service deployable at more sites. Of course other labs may have a file sharing service but again we face the access problem: an external person cannot use it without an account or a token and questions about interoperability need to be considered; furthermore such share-and-sync services usually can provide only modest storage with limited integration, making it mostly applicable to small, final outputs.

\subsection*{Data transformation services}

An important step of an analysis is to transform data from one format to another or to create a custom selection of the data. Examples include: physics groups produce ntuples and then users can further refine the event selections; users developing new ML algorithms and parallelized workflows may want to transform ROOT files to other formats developed outside of HEP for ease of use with other tools; users may need to augment existing formats with custom information. Usually these steps are carried out in isolation, but a logical goal is to simplify and reduce the number of steps required, particularly if they have to be carried out on different resources, i.e. grid first and then local resources and require to register and manage transformed copies and take more storage.  Transformation services, such as ServiceX~\cite{servicex}, have been introduced to give users the ability to transform data on the fly. Such services need to be fully integrated with performant storage and caches to serve the data in a timely manner and could be prime consumers of the object stores mentioned in the previous section. This introduces a dependency on new intermediate services that need to serve multiple users and cache intermediate data for further re-use. The efficiency of these intermediate services are being tested to understand if they are a viable solution. 

\subsection*{Open Questions}

\subsubsection*{On data “locality” at facilities}

Should analysts expect the data, particularly reduced formats, to be local to any facility they wish to use (thus providing low latency access)? Often analysts work with derived datasets (with extra cuts, derived variables), does the same apply? 

\subsubsection*{On POSIX file access?}
Is POSIX required? Maybe just interacting with an object store via e..g XRootD / https / analysis software is sufficient? How much work is required to fully support object stores? Users like filesystem-like semantics, but what part of POSIX is really needed and can we decouple mass storage access from more interactive, smaller scale activities?

\subsubsection*{On user interaction with Distributed Data Management systems}
Can all Distributed Data Management queries be hidden from the user and is this desirable? Should users expect that this is managed for them? 

\subsubsection*{On provenance of intermediate data products}
Which intermediate files need to be promoted from local to global storage so that users can run at different sites and how is this declared by the user? 

\subsubsection*{On common file sharing services}
Do we need a common file sharing service to help users share their files? In which case can any existing services fulfill this purpose?

\section{Federated Identity Management and AAI}
\label{aai}

Equitable access to resources in a VO-wide manner is one of the foundations on which the success of the grid is built. Users can access worldwide resources but any one workload is routed to a convenient resource given, e.g. data constraints. This is based on an agreed key piece of Federated Identity Management (FIM) infrastructure: x509 certificates. So far the majority of end user analysis has not required the use of FIM because the resources used are either private ---  provided by universities or national labs --- with their own AAI mechanisms, or are provided by CERN, which is used by so many users that it is overloaded as a consequence. If there is a significant increase in resources requested, and the need to support all users, FIM will become a key requirement for analysis infrastructure. 

While the WLCG FIM model is very successful, the use and maintenance of certificates for users is difficult and incompatible with the use of cloud services and resources. WLCG is working towards an AAI based on Oauth2.0 and tokens which will replace x509 certificates. By HL-LHC the FIM on the grid will be based on the new token infrastructure, which is fully compatible with cloud and web services such as JupyterHub, CERNBox and kubernetes. This will greatly improve the user experience.  It is highly desirable that the analysis infrastructure services be fully integrated with the new AAI. This integration does not imply or necessitate that all analysis resources support all users, but it allows for a uniform access to all resources for authorized users. As expressed in the user requirements, the VOs ultimately can decide on the authorization mode, i.e. if the facilities will be usable by all users, but whatever decision is made the infrastructure should support it. This would also allow private resources to be more easily made accessible to selected colleagues, which would be a site or local group decision. Some AFs in the US are an example of the latter, currently giving access to ATLAS and CMS users depending on their affiliation.

It is important to underline that currently a Oauth2.0 based AAI is not compatible with a classical SSH/POSIX/batch system infrastructure and development work is necessary to integrate them.

\section{Accelerator  Resources}
\label{accelerators}

The most commonly used accelerators in HEP are GPUs and this is the focus of the discussion here. It is expected that both availability and demand for GPU resources is going to increase, not least in the context of HPC facilities. It is important that GPU resources be made available both for interactive and non-interactive workloads. Particularly in machine learning, a large fraction of time is spent in R\&D to develop a candidate model interactively, before large-scale non-interactive trainings are launched, e.g. for hyperparameter tuning. 

Aside from pure availability of heterogeneous resources, it is important for AFs to provide software hooks to integrate such hardware. WLCG computing has until recently been based on the assumption of relative uniformity of hardware resources. The exclusive use of x86 architecture guaranteed that user code could run everywhere. GPUs, or accelerators in general, pose a new challenge for AFs and WLCG sites. The variety of GPU models, functionality and software needed to access them, as well as the complex memory management, make GPUs more difficult to insert both in a local context as well as at distributed facilities.  The user code is still the subject of R\&D and the requirements, not yet as well defined as for CPUs, vary from GPU sharing, to partial offload of the code to a GPU,  to use of multi-GPUs for large ML models, to requests for specific versions of libraries. How to tune batch systems to support this variety of configurations and possible user requests is still being investigated by sites. For this reason GPUs so far have been used mostly on dedicated resources, such as the HLT farms for very specific workflows, or users have requested single (dedicated) interactive GPUs. In the few cases where multiple GPUs have been requested by a single user they tend to be allocated on a single node. Particularly interesting is the case of interactive GPUs which remain idle while the user is coding, especially when used as a notebook resource that is assigned to the session. Time sharing could be considered in these cases. Physical partitioning is also an option in newer GPU models, and permits providing multiple GPU flavors. Some AFs, like Nebraska and FNAL elastic AF, are deploying specialized software, such as NVIDIA's Triton inference servers, for their inference-as-a-service infrastructure in parallel to their batch system access. The Triton server schedules the model requests and can be used from either local HTCondor jobs or from the interactive nodes.

Users also use commercial cloud resources to access GPUs, but this type of access is usually not integrated. As demand for such resources is still very dynamic, it may be interesting to investigate models of AFs that couple with cloud computing resource providers, where such resources can be allocated on-demand. This model has been used by LHC collaborations, such as ATLAS, to investigate the technical compatibility of collaboration software, as well as validation of the physics output, for other types of non-x86 architectures like ARM and this coupling can also be used for accessing various types of accelerators, more commonly GPUs. CMS, LIGO and protoDUNE have also used commercial providers for integrated inference-as-a-service. In these cases the user interface can either be independent or mediated through the experiment grid WFMS (Workflow Management System). 

\section{Analysis portability and preservation}
\label{preservation}

A core requirement of an AF then is providing the user with tools that help them realize their desired software stack and to address the user portability and preservation requirements expressed in section~\ref{userperspective}. There are few methods users can use to build the software stack.

\subsection*{CVMFS}
CVMFS is the read-only distributed file system for global distribution of HEP software, in which software packages are laid out in a directory structure with software tools that can selectively enable a certain combination of packages within a given user session (e.g., LCG views). The user can augment this global set of software through local installations, which may or may not be carried out in the user space. While this is an effective method to centrally maintain and distribute software it has disadvantages: user code built against CVMFS needs runtime access, so will not work on non-CVMFS nodes; and the user code needs preserved separately from the CVMFS dependencies, which is important for analysis preservation.

\subsection*{Conda}

Conda~\cite{conda} is a cross-platform and language agnostic package and environment manager, which solves portability between collaborators and is adopted particularly when python external tools are used. It is not particularly designed for preservation but it makes it easy for users to build their code in multiple environments with different sets of dependencies and switch from one to another. To be truly preserved the conda environment needs to be locked to specific versions (not the default) and, in addition, the user’s supplementary code needs to be also preserved.

\subsection*{Linux containers}

Linux containers allow users to assemble the software stack on their own, in a self-contained, preservable format. The universal container format that can be read by any runtime is Docker~\cite{docker}, and Docker container registries can be used to distribute images to facilities. If built with fixed dependencies a container does not change and contains all the software it needs to run the application. It should be noted that preserving only the containers may not be sufficient when older versions of OSs ( for example slc5)  do not run anymore on modern hardware --- this remains an area of R\&D. 

Conda, containers and CVMFS can be combined. For example LHCb uses conda in combination with CVMFS to distribute and preserve different user setups~\cite{lbconda}. Since it is a package manager it can also be used to build containers instead of the specific container OS package manager.CVMFS is also used to distribute singularity/apptainer~\cite{singularity} images on the grid --- currently CVMFS stores and serves around 3000 custom apptainer images. This could be extended to other runtimes with containerd~\cite{containerd} images serving facilities that are built on top of cloud-computing infrastructure, such as kubernetes. Supporting different runtime images raises questions about the infrastructure as standard distribution, apptainer images and containerd images would all be duplicated software installations.

The construction of well-maintained software stacks and  images may require expertise from AF and experiment experts and an infrastructure to make it easy for the users. For example, SWAN makes a curated set of software environments available through a dedicated user interface. Projects such as repo2docker~\cite{repotodocker} and Binder abstract the image building process away from the user. Here, the user gives their desired set of requirements in a declarative format which then gets built ad-hoc during a start up phase. For reproducibility and shareability of analyses, users code should be versioned or tagged along with the Git repository and environments (conda or container images) used. This issue was raised at the user sessions conducted within the forum. 

\section{Monitoring and Metrics}
\label{monitoring}

To provide an extensive overview on how resources are used in order to guide infrastructure development and to allow users to make an informed decision about which infrastructure to use, key parameters should be published for each AF.

Metrics for AFs can be categorized into four areas

\begin{enumerate}
    \item User experience metrics
    \item User trend metrics
    \item Performance metrics
    \item Facility metrics
\end{enumerate}

User experience metrics relate to the analysts’ impression of the facility. Support for users is a key component of the analyst experience. Such support metrics include time to acknowledge tickets and, as a separate metric, time to resolve tickets. Unlike time to acknowledge tickets, time to resolve tickets is very case specific, for instance, if a ticket requests or requires the implementation of new functionality it will often take considerable time (and not all requests are even reasonable!). Over an extended time period, and with some care, this can be compared between sites. Once new functionality or features are added to a facility the rate of uptake by users is a useful metric to gauge user engagement and is closely related to the findability and usability of documentation and tutorials, which are key for user on-boarding and retention (see also section~\ref{documentation}). User satisfaction surveys can gather feedback directly from users (albeit response rates can be low). 

User trend metrics can be used to inform future infrastructure development and act as a “live survey” of user analysis requirements. This could include metrics on the data source being used (local or cached from a DataLake), input data format, resources employed per job and interactive session length which would impact parameters of the caching system. In an idealized scenario this would mean a common API across software tools used at the analysis infrastructures, but this cannot realistically be imposed.  Tools like prmon~\cite{prmon}, or in general any code profiler that can be run within the user job, are well tested and can easily collect resource usage information while the jobs run and are not tied to any specific software. Users cannot be expected to run this monitoring on their own, so it requires automation from the VOs and/or the sites. When facilities run these tools, wrapping the jobs automatically, they should also set up a visualization infrastructure for the users to track their jobs. This happens on the grid where the WFMS of the experiments have robust monitoring of every job and metrics are then aggregated to give a wide view of the resource usage by site or globally. Another example of this would be the Grafana dashboard at CERN, which displays some information on the user batch jobs. However, it does not cover other resources (cloud or interactive) or many desirable metrics. Of course tracking basic machine level or batch system resources can be done almost automatically, but tracking each application and knowing what it is, what it does and what data it uses there is no other option but to push the users to use an instrumented framework which collects all this extra information.

The concept of dark analysis was raised at the Analysis Ecosystem Workshop II and refers to the analysis that users run privately on, for instance, university or laboratory resources. We currently have no handle on the type and scale of this analysis and therefore a user survey was proposed to analyze these user trends. This kind of large scale survey must be treated carefully. The survey could be conducted through the HSF Data Analysis working group and would give further insight into the analysis needs of users across experiments. If we had instrumented AF the problem of dark analysis would be reduced if the requisite monitoring was implemented.

Whilst all current facilities have different focuses, and this is important, there are universal aspects of an analysis pipeline that can serve as benchmarks. These can be used to assess analysis facility performance - a performance metric.  The IRIS-HEP Analysis Grand Challenge (AGC)~\cite{agc} has established complete analysis workflows using CERN Open Data that can be used to test the functionality, integration and performance of an AF. These workflows can be factorized to benchmark performance in specific areas, for instance, data delivery. They also serve as useful tutorials for analysts, showcasing complete, realistic end user analyses that exploit the tools discussed in this document as well as the scientific python ecosystem. 

Publication of facility metrics is considered a necessary aspect. Facility metrics quantify the technical performance of the facility rather than the analysis performance. Resource idle time, job failure rate and downtime should all be monitored by the facility host.

Increasingly, sustainability metrics will be assessed, scrutinized and perform an important role in evaluating facilities’ viability, for instance, one could consider the carbon footprint per user/job. As reported at the 26th International Conference on Computing in High Energy \& Nuclear Physics (CHEP 2023) the environmental footprint of HEP research overall can be controlled via modernization of the facilities we use to run jobs, improvements in the software and computing models of the experiments and improvement in hardware technologies. AFs have the opportunity to influence two of these factors. This will be carefully balanced against operational cost per user/job where the weighting of each of these (potentially conflicting) optimisations will be decided by funding agencies. Facilities must provide information for both and meaningful metrics will evolve over time for all sites, not just AFs.

Infrastructure status reports would be important to track the AF efficiency, but it is a whole project in itself and would take a fair amount of operational overhead to maintain, correct, ensure the system reports fairly, etc. Facilities should certainly track metrics internally as a first step. All analysis infrastructures should define key parameters to measure success over time and a level of coordination on these parameters across infrastructures should be implemented to make fair comparisons. As a minimal set we propose the following as an evolution over time: total number of active users, total compute time spent both on local and on batch resources, storage space available and used, I/O rates, Time-To-Completion and data source (local, cached, remote).

\section{End user documentation}
\label{documentation}

The ability for all users, regardless of technical capability, to use and profit from AFs is crucial. The necessary developments in analysis techniques and infrastructure must therefore be undertaken in collaboration with analysts to keep the “barrier-to-entry” on using new analysis facilities/infrastructures as low as possible. Analysts will expect comprehensive documentation and, more importantly, tutorials on how to run analyses that are similar to theirs on the facility - as an example the IRIS-HEP Analysis Grand Challenge has complete analysis workflows, including the configuration of facilities, using open-source code and CERN Open Data that can serve as valuable instructional resources. Tutorials for analysis specific tasks that illustrate how to most effectively use AFs will also be required. These tutorials should be regularly tested on the facilities (perhaps as part of a periodic testing schedule). We can also target experiments’ training and onboarding initiatives such as the LHCb Starterkit and the CMS Data Analysis School with hands-on tutorials to train early career researchers on these infrastructures from the outset --- these initiatives are documented in~\cite{Hageboeck:2023kcb}. The findability, and usability of this documentation and instructional material should be considered an important user metric. It is clear that such educational material needs to rely on a degree of uniformity between the user interfaces --- hiding, as much as possible, the differences between facilities --- to avoid filling the documentation with caveats.

End-user support can be provided both by the facility and by peers. Forums such as Slack/Discord channels provide vital, realtime, peer-to-peer support among countless communities and could be utilized for end-user support on a facility. The facility should endorse such a channel's existence and direct users to it as an official source of support.

\section{Summary}

Motivated by discussions at the HEP Software Foundations (HSF) Analysis Facilities forum, the Analysis Ecosystems II workshop, and the 26th International Conference on Computing in High Energy \& Nuclear Physics, this white paper documents HEP analysis infrastructure goals and the technologies that can be employed to achieve them. These goals should guide the development of such future infrastructures to enable the full scientific potential of HEP experiments in the next decades to be realized. We recommend the formation of cross-experiment working groups to further these goals as a community. In particular to collect the analysis workflows use-cases, tools for monitoring of user jobs and to discuss the integration of object stores into the experiments.


\sloppy
\raggedright
\printbibliography[title={References},heading=bibintoc]

\clearpage

\end{document}